\documentclass[aps,showpacs,longbibliography,notitlepage,superscriptaddress,twocolumn,nofootinbib]{revtex4-1}
\pdfoutput=1
\usepackage[utf8]{inputenc}
\usepackage[english]{babel}
\usepackage[T1]{fontenc}
\usepackage{amsmath}
\usepackage{xcolor}
\colorlet{myPurple}{blue!40!red}
\colorlet{myPurplee}{blue!10!red}
\colorlet{myCyan}{cyan!60!gray}
\colorlet{myRed}{red!66!black}
\colorlet{correction}{black}
\usepackage{tikz}
\usepackage{pgfplots}
\pgfplotsset{compat=1.14}
\usepackage[colorlinks=true,citecolor=myRed,urlcolor=myRed,linkcolor=myRed]{hyperref}
\usepackage[normalem]{ulem}
\usepackage{exscale}
\usepackage{bbm}
\usepackage{graphicx}
\usepackage{amsmath}
\usepackage{latexsym}
\usepackage{amsfonts}
\usepackage{amssymb}
\usepackage{times}
\usepackage[T1]{fontenc}
\usepackage{amsthm}
\usepackage{enumerate}
\usepackage{bbold}
\usepackage{color}
\usepackage{nicefrac}
\newcommand{\sket}[1]{{\ensuremath{\lvert#1\rangle}}}
\newcommand{\lket}[1]{{\ensuremath{\left\lvert#1\right\rangle}}}
\newcommand{\ket}[1]{\if@display\lket{#1}\else\sket{#1}\fi}

\newcommand{\sbra}[1]{{\ensuremath{\langle#1\rvert}}}
\newcommand{\lbra}[1]{{\ensuremath{\left\langle#1\right\rvert}}}
\newcommand{\bra}[1]{\if@display\lbra{#1}\else\sbra{#1}\fi}

\newcommand{\sbraket}[2]{{\ensuremath{\langle#1\rvert#2\rangle}}}
\newcommand{\lbraket}[2]{{\ensuremath{\left\langle#1\!\left\rvert\vphantom{#1}#2\right.\!\right\rangle}}}
\newcommand{\braket}[2]{\if@display\lbraket{#1}{#2}\else\sbraket{#1}{#2}\fi}

\newcommand{\sketbra}[2]{{\ensuremath{\lvert #1\rangle\!\langle #2\rvert}}}
\newcommand{\lketbra}[2]{{\ensuremath{\left\lvert #1\right\rangle\!\!\left\langle #2\right\rvert}}}
\newcommand{\ketbra}[2]{\if@display\lketbra{#1}{#2}\else\sketbra{#1}{#2}\fi}
\usepackage{tcolorbox}
\usepackage{mathtools}

\usepackage{tikz}
\usepackage{lipsum}
\theoremstyle{plain}

\usepackage{graphicx}
\usepackage{bm}
\usepackage{dsfont}
\usepackage{tikz}
\usepackage[T1]{fontenc}
\usepackage{amsthm}
\usepackage{array}
\usepackage{amssymb}
\usepackage{amsfonts}
\usepackage{cancel}
\usepackage[toc,page]{appendix}
\usepackage{multirow}
\usepackage{color}
\usepackage{calrsfs}

\usetikzlibrary{backgrounds,decorations.pathreplacing,calc}

\usepackage{tkz-euclide}

\DeclareMathAlphabet{\mathcal}{OMS}{cmsy}{m}{n}

\newtheorem*{theorem*}{Theorem}

\begin{document}

\author{Antoine Girardin}
\affiliation{Département de Physique Appliquée, Université de Genève, 1211 Genève, Switzerland}
\author{Nicolas Brunner}
\affiliation{Département de Physique Appliquée, Université de Genève, 1211 Genève, Switzerland}
\author{Tamás Kriváchy}
\affiliation{Département de Physique Appliquée, Université de Genève, 1211 Genève, Switzerland}
\affiliation{Institute for Quantum Optics and Quantum Information — IQOQI Vienna,
Austrian Academy of Sciences, Boltzmanngasse 3, 1090 Vienna, Austria}
\affiliation{Atominstitut, Technische Universit\"at Wien, 1020 Vienna, Austria}

\title{Building separable approximations for quantum states via neural networks}

\date{\today}

\begin{abstract}
Finding the closest separable state to a given target state is a notoriously difficult task, even more difficult than deciding whether a state is entangled or separable. To tackle this task, we parametrize separable states with a neural network and train it to minimize the distance to a given target state, with respect to a differentiable distance, such as the trace distance or Hilbert--Schmidt distance. By examining the output of the algorithm, we  \textcolor{correction}{obtain an upper bound on the entanglement of the target state}, and construct an approximation for its closest separable state. We benchmark the method on a variety of well-known classes of bipartite states and find excellent agreement, even up to local dimension of $d=10$, \textcolor{correction}{while providing conjectures and analytic insight for isotropic and Werner states}. Moreover, we show our method to be efficient in the multipartite case, considering different notions of separability. Examining three and four-party GHZ and W states we recover known bounds and obtain additional ones, for instance for triseparability. 
\end{abstract}

\maketitle

\section{Introduction}
Entanglement is now considered a defining feature of quantum theory, with broad implications in modern physics, from quantum information processing to many-body physics.

The detection and characterisation of entanglement is however a notoriously challenging problem \cite{horodecki_quantum_2009,guhne_entanglement_2009}. First of all, it is known that the problem of determining whether a given density matrix is entangled or separable is NP-hard \cite{gurvits_classical_2003, gharibian_strong_2010}. There exist however general methods for detecting entanglement, notably the celebrated negativity under partial transposition (NPT) criteria which ensures the considered density matrix must be entangled \cite{peres_separability_1996,horodecki_separability_1996}. The converse, however, does not hold, as there exist entangled states which are positive under partial transposition, so-called bound (or PPT) entanglement~\cite{horodecki_mixed-state_1998}. Other techniques have been developed, yet all of them are only useful in specific cases in practice. In particular, Ref.~\cite{navascues_complete_2009} developed a method based on semi-definite programming, while Refs~\cite{barreiro_experimental_2010,kampermann_algorithm_2012} proposed a numerical algorithm to construct separable decompositions. Moving beyond the bipartite case, the certification of multipartite entanglement, of which there exist a zoology of different forms, is by far even more challenging and less understood.

Beyond the question of determining whether a given quantum state is entangled or not, one may consider the problem of approximating a given target state via a separable one. More precisely, if the target state is separable, the question is to provide an explicit (separable) decomposition for the density matrix. While, if the state is entangled, to construct a separable state that minimizes a certain distance (in the Hilbert space) with respect to the target state. 


This question has been addressed indirectly in the studies of entanglement measures based on the distance from the set of separable states \cite{horodecki_quantum_2009,vedral_quantifying_1997}, and is particularly relevant when constructing entanglement witnesses~\cite{pittenger_convexity_2002,bertlmann_geometric_2002,pittenger_geometry_2003,bertlmann_optimal_2005,bertlmann_geometric_2008}. Additionally, finding the closest separable state has been studied directly, but this task is even difficult for two-qubit systems~\cite{kim_difficulties_2010}. For a very specific notion of distance, it has also been studied directly though the concept of ``best separable approximation'' of a quantum state~\cite{lewenstein_separability_1998}. 
The construction of separable approximations for multipartite states is largely unexplored, except for specific families of states, which typically have a high level of symmetry~\cite{ishizaka_reduction_2002,hayashi_entanglement_2008,hayashi_geometric_2009,hubener_geometric_2009,parashar_entanglement_2011,carrington_geometric_2015,quesada_best_2014,akulin_essentially_2015,rodriques_bounding_2014}. Moreover, Ref.~\cite{shang_convex_2018} developed a numerical method based on Gilbert's algorithm for constructing closest separable approximations for multipartite states, considering various notions of separability.

\begin{figure}[t]
    \centering
    \includegraphics[width =0.48\textwidth]{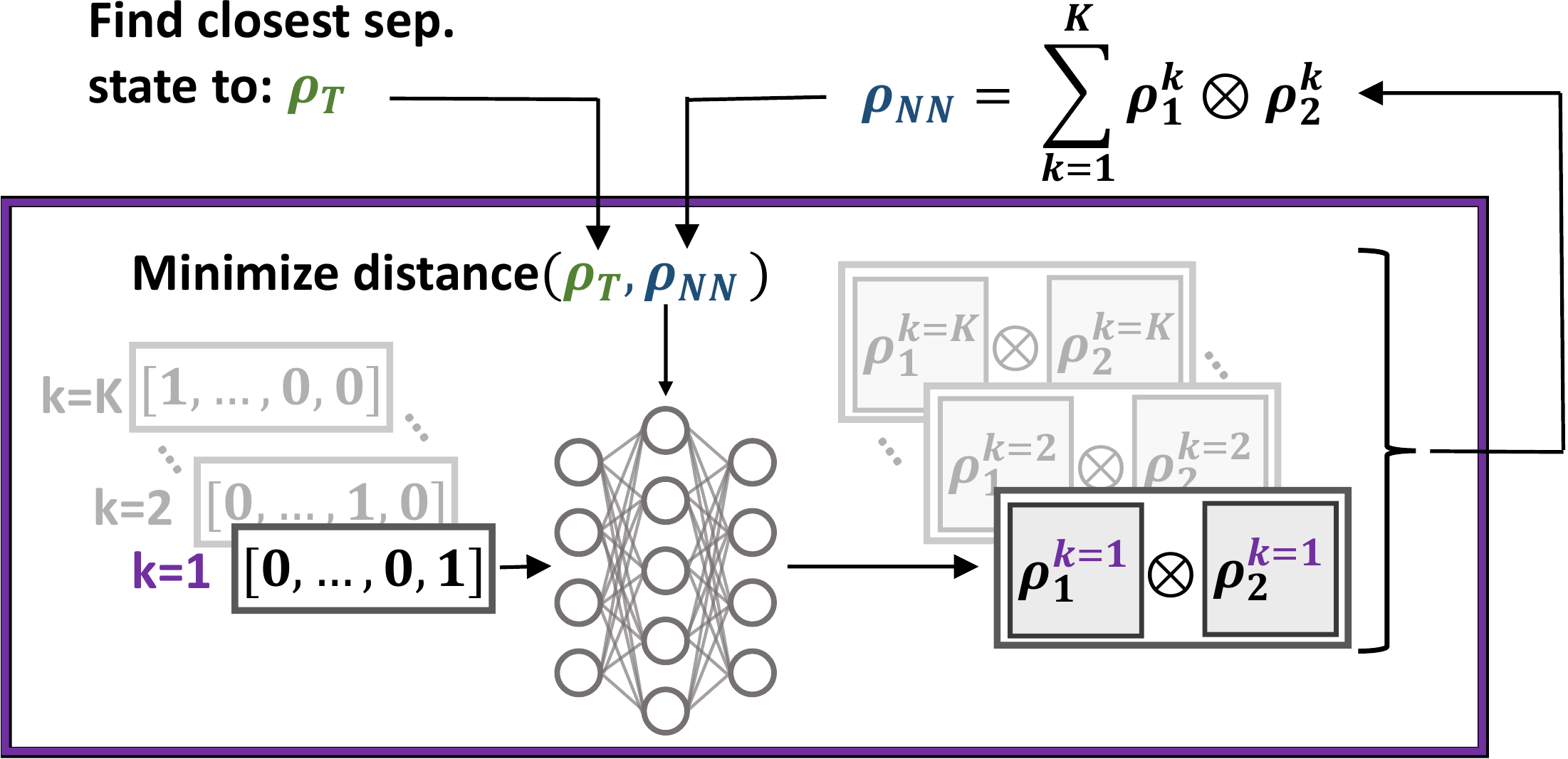}
    \caption{Schematic of the proposed algorithm. Given a target state $\rho_\text{T}$, a neural network constructs a separable state $\rho_\text{NN}$ which minimizes the distance to the target, i.e. it tries to find the closest separable state. For a single input $k$ (represented in the one-hot encoding), the neural network outputs the (subnormalized) pure state $\rho_1^k \otimes \rho_2^k$. The neural network is evaluated for $K$ values of $k$, and its outputs are summed up to construct $\rho_{\text{NN}}$. The distance between this and $\rho_\text{T}$ is used to update the neural network's parameters.}
    \label{fig:schematic}
\end{figure}

In the present work, we attack these questions using tools from machine learning. Specifically, we devise neural networks for constructing a separable approximation, given a target density matrix. We define a notion of ``closest separable state'', which represents the separable state minimizing a given distance with respect to the target; note that this does not coincide with the best separable approximation in general. We benchmark our method with two distance measures, the trace distance and Hilbert-Schmidt distance, on several examples, including bipartite entangled state of local dimension up to $d=10$ (isotropic and Werner states). \textcolor{correction}{From the output of the algorithm, we obtain some analytical insight on the distance to the closest separable state as well as its structure. We also consider a family of states featuring bound entanglement. In turn, we demonstrate the potential of our method in the multipartite case, where we construct multi-separable decompositions for several classes of entangled states (noisy GHZ and W states) up to four qubits. Again, we show how to obtain some analytical bounds on the noise thresholds from the output of the algorithm.} 
In particular, we establish estimates on multi-separability.
We conclude with a number of open questions and directions for future research. \textcolor{correction}{Finally, in the Appendices we study the case of randomly chosen two-qubit states, for which we create ans\"atze for closest separable states, and derive an exact bound for the two-qubit case.}

\section{Related work}

Previous work on using machine learning for the separability problem has been focused either having the machine choose good measurements and then using an existing entanglement criteria~\cite{wang_learning_2017,yosefpor_finding_2020} , or on viewing the task as a classification problem~\cite{lu_separability-entanglement_2018,gao_experimental_2018,ma_transforming_2018,gray_machine-learning-assisted_2018,yang_experimental_2019,goes_automated_2021,ren_steerability_2019}. For classification, typically a training set is constructed where quantum states are labeled as separable or entangled. The machine learns on this training set and given a new example predicts whether it is entangled or separable. There are several difficulties with this approach. First, the machine just gives a guess of whether the state is entangled or separable, and does not provide any kind of certificate. Second, the training data can only be generated in a regime where we already understand the problem well, which results in the machine giving only marginal new insight at best. This could be circumvented by using suboptimal criteria (e.g. PPT) to create the training data, however, the machine would just learn this criteria instead of correctly identifying the entanglement and/or separability boundary.

We overcome these challenges by using a generative model, which tries to give an explicit separable decomposition of a target state. This way, we immediately get a certified upper bound on the distance from the separable states. A similar approach has been taken in Refs.~\cite{harney_entanglement_2020,harney_mixed_2021}, where the authors represent the quantum states with ``quantum neural network states''~\cite{carleo_solving_2017,melko_restricted_2019}, and their extension to density matrices~\cite{yoshioka_constructing_2019,hartmann_neural-network_2019,nagy_variational_2019,vicentini_variational_2019}, as opposed to the dense representation we utilise. Their results show a more limited flexibility in the loss function and in the design of types of separable states\textcolor{correction}{, in particular in the multipartite case. In contrast, our technique allows us to examine the key notion of genuine multipartite entanglement \cite{horodecki_quantum_2009,guhne_entanglement_2009}, the strongest form of entanglement in multipartite quantum systems. This is possible as our model allows for optimising over biseparable models. Hence we can obtain close to optimal bounds for demonstrating 3-party and 4-party genuine multipartite entanglement. Moreover, our method also allows us to tackle the even harder question of triseparability.}    

\section{Preliminaries}
In this section we first introduce the notions of separability for bipartite and multipartite systems and then define the closest separable state. Finally we introduce the basic concepts of neural networks. For more detailed introductions on separability and entanglement or on neural networks, we refer the interested reader to Refs.~\cite{horodecki_quantum_2009} and \cite{goodfellow_deep_2016}, respectively.

A quantum state $\rho_{12}$ acting on $\mathcal{H}_1\otimes\mathcal{H}_2$, shared between two parties, is said to be \emph{separable} if it can be constructed by the convex combination of some local quantum density matrices $\rho_1^k$ acting on $\mathcal{H}_1$, and $\rho_2^k$ acting on $\mathcal{H}_2$ as
\begin{align}\label{eq:separable}
\rho = \sum_{k=1}^K p_k \rho_1^k \otimes \rho_2^k,
\end{align}
with $\{p_k\}_k$ a normalized discrete probability distribution. Any state which is not separable is \emph{entangled}. For finite dimensional systems, i.e. where $d_i:=\text{dim} (\mathcal{H}_i) < \infty$, for $i\in\{1,2\}$, the local states of the decomposition, $\rho_i^k$, can be taken to be pure. Due to Caratheodory's theorem, the number of terms required in the sum, $K$, is upper bounded by $(d_1 d_2 )^2$.



For a multipartite system of $n$ parties several notions of separability exist. The straightforward generalization of Eq.~(\ref{eq:separable}) results in the notion of a \emph{fully separable} decomposition,
\begin{align}\label{eq:fully_separable}
\rho = \sum_{k=1}^K p_k \rho_1^k \otimes \rho_2^k \otimes \dots \otimes \rho_n^k.
\end{align}
Naturally, one can also just examine bipartite separability on the mutlipartite system by grouping the parties together. This leads to the notion of \emph{biseparability with respect to the partition $(\mathcal{I}|\mathcal{\bar{I}})$},
\begin{align}\label{eq:biseparable_given_partition}
\rho = \sum_{k=1}^K p_k \rho_{\mathcal{I}}^k \otimes \rho_{\bar{\mathcal{I}}}^k,
\end{align}
where $\mathcal{I}$ denotes a subset of the indices $\{1,2,\dots,n\}$ and $\bar{\mathcal{I}}$ denotes its complement. A multipartite state is called \emph{biseparable} if it can be decomposed as a convex mixture of states that are separable considering all possible bipartitions, namely
\begin{align}\label{eq:biseparable}
\rho = \sum_{k=1}^K p_k \rho_{\mathcal{I}_k}^k \otimes \rho_{\bar{\mathcal{I}_k}}^k,
\end{align}
where crucially, now each $I_k$ can be different.

There are many ways to quantify entanglement of a target state $\rho_\text{T}$, among which a particularly useful one is based on the distance of a state from the set of separable states. Any distance measure\footnote{We use the term distance, in line with the literature, however note that $D$ must not necessarily be a metric, and is thus more related to the notion of a divergence.} $D$ between quantum states $\sigma_1, \sigma_2$, which is zero if and only if $\sigma_1=\sigma_2$, and for which $D(\sigma_1,\sigma_2) \geq D\left(\Lambda(\sigma_1),\Lambda(\sigma_2)\right)$ for any completely positive trace preserving map $\Lambda$, can be used to construct an entanglement measure, by minimizing $D(\rho_\text{T},\rho_\text{Sep.})$ over separable states $\rho_\text{Sep.}$ \cite{horodecki_quantum_2009,vedral_quantifying_1997}.
We will use the neural network to find the \emph{closest separable state} with respect to a distance $D$, formally
\begin{equation}
\rho_\text{CSS} :=  \underset{\rho_\text{Sep}}{\arg \min}\, D(\rho_\text{T},\rho_\text{Sep.})),
\end{equation}
where $\rho_\text{Sep}$ is a separable state. Note that the closest separable state is not necessarily unique. For the neural network method presented in this paper, any $D$ which is differentiable with respect to one of the states can be used. We choose to work with two distances; the first is the trace distance (related to the Schatten 1-norm)~\cite{eisert_remarks_2003},
\begin{equation}
D_{\text{Tr}}(\sigma_1,\sigma_2) = \frac{1}{2} \text{Tr}\sqrt{(\sigma_1-\sigma_2)^2}= \frac{1}{2} \sum_i |\mu_i|,
\end{equation}
where $\{\mu_i\}_i$ are the eigenvalues of $\sigma_1-\sigma_2$. Note that the trace-distance-based measure can be useful in quantum hypothesis testing, and, among other measures, is an important measure in the study of closest \emph{classical} states, which is distinct from the closest separable state~\cite{aaronson_hierarchy_2013,paula_geometric_2013,nakano_negativity_2013,modi_unified_2010,bellomo_unified_2012}. We will not examine closest classical states in this work, but note that our methods can easily be adopted for their study.

The second distance we consider is the Hilbert-Schmidt distance (related to the Schatten 2-norm)~\cite{vedral_entanglement_1998,witte_new_1999,krammer_characterizing_2009}
\begin{align}
D_{\text{HS}}(\sigma_1,\sigma_2) &= \sqrt{\text{Tr} \left[ (\sigma_1-\sigma_2)^2\right]}.
\end{align}
The Hilbert--Schmidt-based measure can be useful for constructing entanglement witnesses~\cite{pittenger_convexity_2002,bertlmann_geometric_2002,pittenger_geometry_2003,bertlmann_optimal_2005,bertlmann_geometric_2008}. Both the trace distance and Hilbert-Schmidt distance can be used as a basis for an entanglement measure, however, one could consider others, such as the Bures distance~\cite{vedral_entanglement_1998}, relative entropy of entanglement~\cite{vedral_quantifying_1997} or the robustness of entanglement~\cite{vidal_robustness_1999}; see e.g. Ref.~\cite{zyczkowski_introduction_2006} for an overview and other examples of geometric measures of entanglement.

Let us now concisely introduce the concept of an artificial neural network~\cite{goodfellow_deep_2016}, the basis of our numerical representation of separable states. A neural network is a numeric model which can in principle represent any multivariate function. A crucial point is to be able to adjust the parameters of the neural network to represent the desired function, however in many use-cases this can be done surprisingly efficiently with the techniques of deep learning.

In this work we will be using one of the simplest types of neural networks, the so-called multilayer perceptron. It is characterized by the number of neurons per layer (width), the number of layers (depth), and the activation functions used at the neurons. Altogether these model an iterative sequence of \emph{parametrized} affine, and \emph{fixed} nonlinear transformations, on the input; namely the map from layer $l$ to $l+1$ is
\begin{align}
r_{l+1} = h(W_l r_l + b_l),
\end{align}
where the weight matrix $W_l$ and bias vector $b_l$ parametrize the affine transformation, $h$ is a fixed differentiable nonlinear function (activation function), and $r_l$ is the input of layer $l$, and its length signifies the width (number of ``neurons'') of layer $l$. The vector $r_1$ ($r_{\text{depth}}$) is the input (output) of the whole model. At initialization, the weights and biases of all layers are set randomly. During training, the parameters of the model ($\{W_l,b_l\}_l$) are updated such that they minimize a differentiable loss function of the training set, which as we will see later, in our case will be the trace or Hilbert--Schmidt distance. This is done by first evaluating the model for a \emph{batch} of inputs, and then by slightly updating the parameters via a method called backpropagation, which relies on the gradient of the loss function with respect to the model parameters. This is repeated for many batches, until the model converges, a maximum training time is reached, or a satisfactory loss is achieved. Once trained, the neural network can be evaluated on new input instances.

\section{Neural networks as separable states}
The task is to find the closest separable state to a given target density matrix. The central idea of this work is to use a neural network as a variational ansatz for the density matrix by representing the local components of the separable decomposition with a single neural network. The approach is inspired by a similar approach taken for nonlocality, where neural networks represent the local components of a Bell-local behavior~\cite{krivachy_neural_2020}.

To demonstrate the method, let us examine the example of a bipartite 2-qubit state. We ask a neural network to represent the map
\begin{equation}\label{eq:parametrized_bipartite}
k \rightarrow (p_k, \rho^k_1, \rho^k_2),
\end{equation}
where we take $\rho^k_i=\ket{\psi^k_i}\bra{\psi^k_i}$ to be pure states, with $\ket{\psi^k_i}=  \alpha^k_i\ket{0}+\beta^k_i\ket{1}$. That is, the neural network will take as input an integer value $k$ between $1$ and $K$ (in a one-hot representation), and will output the numbers $(p_k, \alpha^k_1, \beta^k_1, \alpha^k_2, \beta^k_2)$, such that normalization for each subsystem is satisfied. Note that for each complex number, two real numbers are output, the real and imaginary part. We evaluate the neural network for $K$ values of $k$, normalize the $(p_k)_k$ probability vector and sum up the outputs $p_k \rho^k_1 \otimes \rho^k_2$ to construct a separable state via Eq.~(\ref{eq:separable}), namely $\rho_{\text{NN}}=\sum_{k=1}^K p_k \rho^k_1 \otimes \rho^k_2$. The neural network is trained to minimize the distance between the target density matrix $\rho_\text{T}$ and the constructed separable density matrix $\rho_\text{NN}$, i.e. $D_{\text{Tr}}(\rho_\text{T},\rho_\text{NN})$. The process is roughly illustrated in Fig.~\ref{fig:schematic}, where the $p_k$ are not shown explicitly.

By construction the neural network represents a single density matrix $\rho_{\text{NN}}$, so for each target state $\rho_\text{T}$, the network must be retrained in order to obtain an approximation of the closest separable state to that target state. During training, requiring $K$ values of $k$ in order to evaluate the state technically means working with a batch size of size $K$. That is, we evaluate $K$ inputs ($k=1,2,\dots K$) to construct $\rho_\text{NN}$ and only then calculate the gradients required for the optimization of the neural network. \textcolor{correction}{An advantage of this method is that only the input layer increases with $K$, which implies that the global size of the neural network can grow slowly with the dimension of the target state.}

More generally, for more parties or higher dimensions, the neural network represents the map
\begin{equation}
k \rightarrow (p_k, \rho^k_1, \rho^k_2, \dots \rho^k_n ),
\end{equation}
where we take the $\rho^k_i$ ($i=1\dots n$) to be pure, and the neural network explicitly outputs the parameters of the pure states. By evaluating this neural network for $K$ values of $k$, we construct a separable state via either Eq.~(\ref{eq:separable}) for the bipartite case ($n=2$), or any of Eqs.~(\ref{eq:fully_separable},\ref{eq:biseparable_given_partition},\ref{eq:biseparable}) for the different notions of multipartite separability. Recall that by Caratheodory's theorem, in principle the largest $K$ needed is $\prod_j d_j^2$, however even less could be sufficient. Thus we keep $K$ as a free hyperparameter, which we set before training begins. More technical details on the neural networks we used can be found in Appendix~\ref{app:NN_details} or in the sample code provided in the Code Availability section.


The neural network is optimized in the high-dimensional non-convex landscape of the network's weights, so it is not guaranteed to converge to the optimal solution. However, in practice, optimization procedures based on gradient descent reach close-to-optimal solutions efficiently. Notice, that even for suboptimal solutions we obtain an upper bound on the amount of entanglement of the target state, since the utilized distances serve as entanglement measures~\cite{vedral_quantifying_1997,vedral_entanglement_1998,horodecki_operational_2000}. However, we can go one step further, and examine families of states parametrized by a single parameter, which we refer to as $q$, typically of the form
\begin{align}
    \rho_\text{T}(q) = q \rho_\text{ent} + (1-q) \rho_\text{sep},
\end{align}
where $\rho_\text{ent}$ is an entangled state and $\rho_\text{sep}$ is a separable state, oftentimes the maximally mixed state. If $\rho_\text{ent} \equiv \rho_\text{T}(q=1)$ is truly entangled, then when decreasing $q$, for some value $q^*$ we will cross the separability boundary. We can observe this transition by varying $q$ and retraining the neural network from scratch for each target distribution. An approximation of $q^*$ becomes clear from how close the algorithm can get to the target states for different $q$ values.

\section{Case studies}

To benchmark the method, we first use the algorithm to examine the separability boundary for some exemplary families of bipartite states. \textcolor{correction}{We consider symmetric classes of states, i.e. isotropic and Werner states, and estimate the noise threshold for separability. We find excellent agreement with analytically known optimal thresholds. Additionally, we compare our results for the minimal Hilbert-Schmidt distance for isotropic states to the known analytic distance, and find excellent correspondence. We furthermore conjecture the minimal trace distances for isotropic, and trace and Hilbert-Schmidt distances for Werner states. 
Moreover, we obtain analytical insight to the problem of characterising the closest separable states for the Bell state in terms of trace distance. Then we also discuss an example of bound entanglement, i.e. an entangled sate that cannot be detected by the partial transpose criterion.}

\textcolor{correction}{
Then we move on to the multipartite case, where we consider 3- and 4-qubit GHZ and W states. We show that our algorithm can capture various notions of multipartite entanglement, including full separability, biseparability and even triseparability. We estimate again noise thresholds, finding excellent agreement with previously known bounds. Moreover, we explain how, from the numerical output from the algorithm, one can obtain analytical bounds on the noise thresholds.}

Additionally, in Appendix~\ref{app:SGD} we compare our neural network algorithm to a naive gradient-descent based heuristic to show its advantage. 
In Appendix~\ref{app:random} we examine the performance of the algorithm on random bipartite two-qubit density matrices, \textcolor{correction}{ and compare it to the optimal solution obtained via a semi definite program (SDP).} We conjecture an analytic ansatz of the closest separable state for a general 2-qubit state, which we find to be very close to the solutions found by the neural network, and prove a bound on the trace distance. \textcolor{correction}{In Appendix~\ref{app:thresholds}, we detail a method to obtain a strict lower bound on the separability threshold compatible with our method.}

\subsection{Bipartite case}

\begin{figure*}[t]
    \centering
    \includegraphics[width =\textwidth]{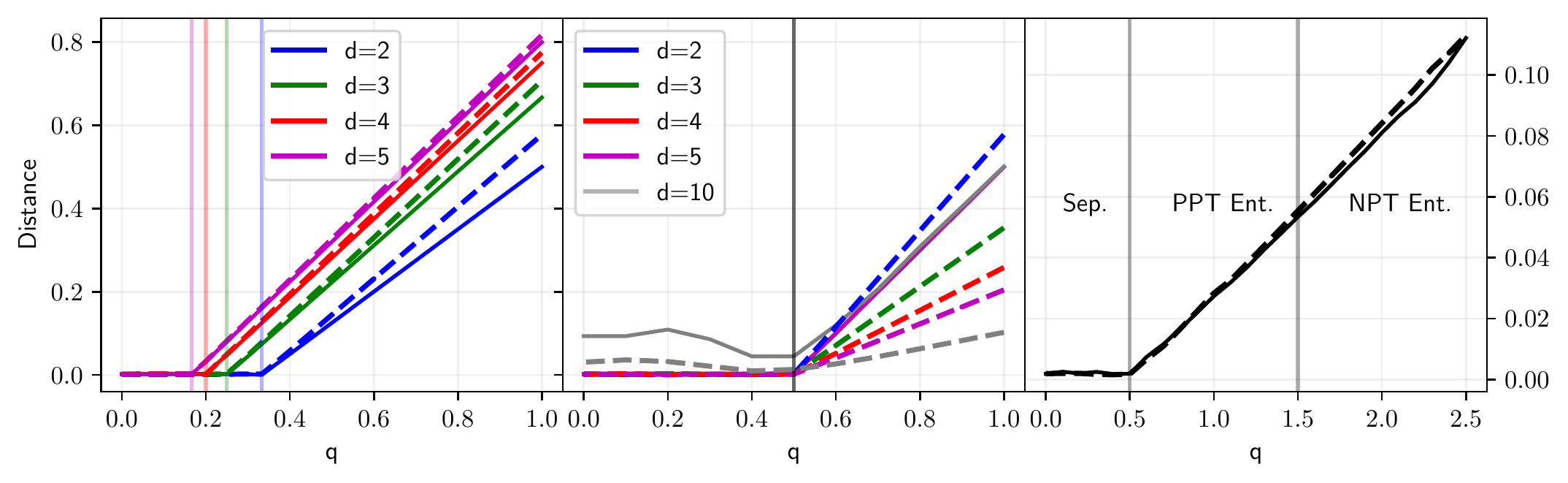}
    \caption{Distance from target state to closest separable state found by the neural network, trained and evaluated on the trace distance (solid lines) or Hilbert--Schmidt distance (dashed lines), trained independently for different values of $q$. Separability thresholds are depicted with vertical lines. Results are shown for $d\times d$ isotropic (left) and Werner (center) states, as well as for PPT entangled states of the form (\ref{eq:minter}) (right). Note that for the Werner states with $2\leq d\leq5$, all the trace distance curves essentially overlap and are thus not distinguishable on the plot. For the right plot we did 2 runs and kept the best results to improve the smoothness of the curves.}
    \label{fig:iso_werner}
\end{figure*}

We start our benchmarking with 
two classes of highly symmetric  bipartite states. 
Isotropic states are defined as
\begin{align}\label{eq:isotropic}
\rho_{\text{iso}}(q) = \frac{1-q}{d^2}\mathbb{I}_{12} + q\ket{\phi^+_d}\bra{\phi^+_d},
\end{align}
where $0\leq q \leq 1$, $\mathbb{I}_{12}$ is the identity operator on the joint space, and 
\begin{align}\label{eq:max_ent_state}
\ket{\phi^+_d} = \frac{1}{\sqrt{d}}\sum_{i=1}^d \ket{i}_1\ket{i}_2,
\end{align}
is the maximally entangled state of local dimension $d$.
Werner states are defined as
\begin{align}
\rho_\text{Werner}(q) = (1-q)\frac{2}{d(d+1)}P_{\text{sym}} + q\frac{2}{d(d-1)} P_{\text{as}},
\end{align}
where
\begin{align}\nonumber
P_{\text{sym}}=& \frac{1}{2}( \mathbb{I}_{12} + F_{12}),\\
P_{\text{as}} =& \frac{1}{2}( \mathbb{I}_{12} - F_{12}),
\nonumber
\end{align}
with $F_{12} = \sum_{i,j}^d \ket{i}\bra{j}_1 \otimes \ket{j}\bra{i}_2$ the flip operator.

These classes of states represent a good benchmark as the separability thresholds, i.e. the value of $q$  for which the state becomes separable, are known analytically. Specifically, 
isotropic states are separable for $q\leq \frac{1}{d}$, while Werner states are separable for $q\leq \frac{1}{2}$. Hence when running our algorithm for these classes of target states, for different values of $q$, we expect to find a distance of the closest separable state that vanishes when $q$ approaches the separability threshold. This is precisely what we observe. We run the neural network independently for 11 values of $q$, and additionally for the exact separability boundary value. The results for both the trace distance and Hilbert-Schmidt distance for $d\leq 5$ are depicted in Fig.~\ref{fig:iso_werner} (each line is plotted with its respective loss function, the trace or Hilbert-Schmidt distance). They confirm that the algorithm works properly in this regime, finding a sharp transition at the known separability thresholds. When making a linear fit to the data that is outside the seemingly flat separable region, we recover the thresholds with a precision of at least $10^{-4}$. To give an example of the running time on a personal computer, for isotropic states the training for a single target state for $d=5$ took at most 15 minutes, while for $d=2$ it took only at most 30 seconds\footnote{Timed with an Intel i7-8700k CPU @ 3.70 GHz with 6 cores (12 threads) and 16 GB RAM.}. When the trace distance is found to be smaller than $2\cdot10^{-3}$, we choose to stop the training, and conclude that the state to be separable. Otherwise we run the algorithm until the resulting trace distance converges, i.e. it doesn't change more than $2\cdot10^{-4}$ in one epoch.

Additionally, for $d=10$ we examine the Werner states, also plotted in Fig.~\ref{fig:iso_werner}. For such a large state, with $K=100$, training took about 1 hours 15 minutes on a personal computer for a single epoch (3000 batches), which was reduced to 45 minutes when training on a GPU\footnote{Trained on a RTX-3080 GPU with 10 GB memory.}. Due to the increased runtime we only ran one epoch for each point in Fig.~\ref{fig:iso_werner}, and did not wait until convergence. We observe that the neural network struggles more in finding a closest separable state in the separable area, however it works remarkably well in the entangled regime, and still manages to give qualitatively interpretable results on where the entanglement boundary lies. For increased accuracy one could run the algorithm several times independently and take the smallest value for each $q$, or one could run the algorithm with a larger batch size $K$. For example for the separability boundary at $q=0.5$, by using $K=150$ instead of 100, after 5 epochs (5 times 3000 batches), the trace distance reduced to 0.024 from the 0.045 seen in Fig.~\ref{fig:iso_werner}.

\begin{figure}[t]
    \centering
    \includegraphics[width =0.5\textwidth]{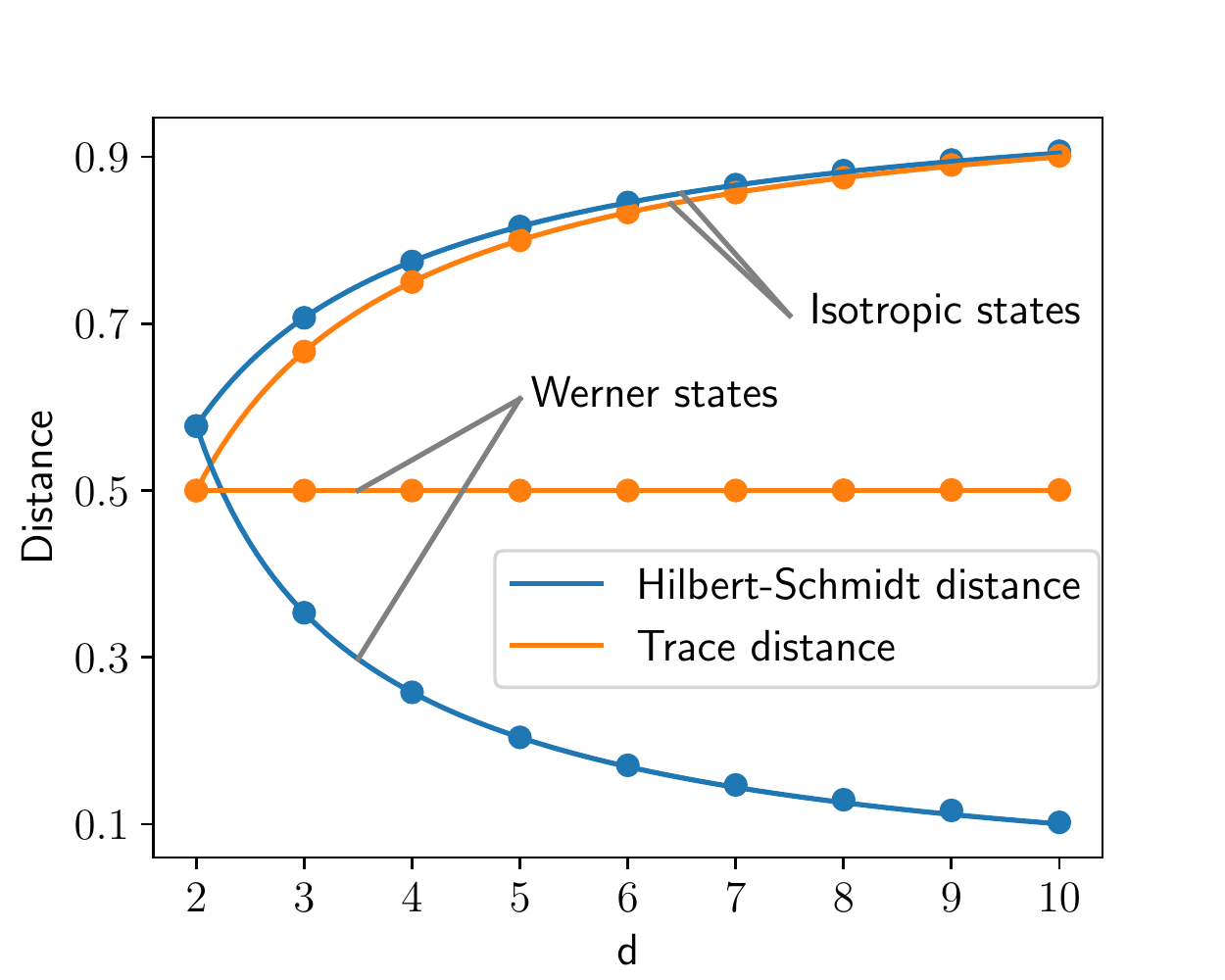}
	\caption{\textcolor{correction}{Distance to the closest separable state for the maximally entangled state $\ket{\phi^+_d}$ (i.e. an isotropic state with $q=1$) and the Werner state as a function of the local dimension $d$. The dots indicate the values found by the neural network. For the Hilbert-Schmidt distance (blue curve), Eq.~(\ref{dhs}) gives the exact expression for the isotropic state as proven in Ref.~\cite{bertlmann_optimal_2005}. We conjecture the general formula for the trace distance of the isotropic state in Eq.~(\ref{dtd}). For the Werner states, we conjecture the general formula for the trace distance in Eq.~(\ref{dtd_wer}) and for the Hilbert-Schmidt distance in Eq.~(\ref{dhs_wer}).}}
	\label{imgiso}
\end{figure}

\textcolor{correction}{
For isotropic states, the exact Hilbert-Schmidt distance to the closest separable state is known. In dimension 2, we retrieve the value of $1/ \sqrt{3}$ that was known shown in Refs.~\cite{witte_new_1999, bertlmann_geometric_2002}. A generalized formula is given by Ref.~\cite{bertlmann_optimal_2005},
	\begin{equation}
		D_{\text{HS}}^\text{iso} = \frac{\sqrt{d^2-1}}{d} \left(q-\frac{1}{d+1}\right),
		\label{dhs}
	\end{equation}
which we recover up to excellent precision for $q=1$ and up to $d=10$ in Fig.~\ref{imgiso}.} 

\textcolor{correction}{
Moreover, we examine the trace distance, and based on the results in Figs.~\ref{fig:iso_werner} and \ref{imgiso}, we conjecture that the trace distance from the isotropic states to their respective closest separable states follows
	\begin{equation}
		D_{\text{Tr}}^\text{iso} = \frac{d^2-1}{d^2}\left(q-\frac{1}{d+1}\right).
		\label{dtd}
	\end{equation}
Using the semidefinite program described in Appendix~\ref{app:random}, we can verify that a trace distance of $1/2$ is optimal for the Bell state, i.e. for $d=2$. We have not found this explicitly proven in the literature, even though the related concept of finding the closest \emph{classical} state to the Bell state has been well studied \cite{paula_geometric_2013,nakano_negativity_2013,aaronson_hierarchy_2013}.}

\textcolor{correction}{
For Werner states, similarly drawing from Fig.~\ref{fig:iso_werner}, we conjecture that the distance to the closest separable is given by 
	\begin{equation}
		D_{\text{HS}}^\text{Wer} = \frac{2}{\sqrt{d^2-1}} \left(q-\frac{1}{2}\right)
		\label{dtd_wer}
	\end{equation}
	for the Hilbert-Schmidt distance, and by
	\begin{equation}
		D_{\text{Tr}}^\text{Wer} = q-\frac{1}{2} \,,
		\label{dhs_wer}
	\end{equation}
for the trace distance. To the best of our knowledge, these relations are not proven in the literature. We note that plotting these conjectured equations in Fig.~\ref{fig:iso_werner} (first two panels) would result in lines that are indistinguishable from the data.}

\textcolor{correction}{While the algorithm performs a numerical optimisation, one can nevertheless obtain some analytical insight from the output. Here we illustrate this point. First we could partially characterize the closest separable state for the two-qubit Bell state, i.e. $\ket{\phi^+_2}$. Setting the Bell state as the target, and using the trace distance as the loss function of the neural network, at the end of training it finds the separable state
\begin{equation}\label{eq:app_NN_ansatz}
\begin{pmatrix}
    (\frac{1}{2}-a)	& c & \bar{c} & a  \\
    \bar{c}			& a & b & -\bar{c} \\
	c			 & \bar{b} & a & -c  \\
    \bar{a}	& -c & -\bar{c} & (\frac{1}{2}-a)
\end{pmatrix},
\end{equation}
with $a=\frac{1}{4}$, $|b|\in[0,1/4]$ and small $c$ values. However, when using the Hilbert--Schmidt distance $D_\text{HS}$ as the loss, the neural network converges to the $a=\frac{1}{6}$, $b=0$ and $c=0$ solution. Both solutions have the same trace distance of $0.5$ from the Bell state. From these two extremes, we constructed the ansatz Eq.~\ref{eq:app_NN_ansatz} for the closest separable state and verify that for $c=0$, $a\in[0,\frac{1}{6}]$ and $b\in[0,a]$ they indeed all give a trace distance of $\frac{1}{2}$. We go further and find other values of $(a,b,c)$ for which the trace distance is $\frac{1}{2}$. For example if all parameters are set to be real, and $b=a$, then it is a closest separable state for $\sqrt{(a-\frac{1}{8})^2 + c^2} \leq \frac{1}{8}$. The same holds for $a=1/6$, $24 c^2 - \frac{1}{6} \leq b \leq \frac{1}{6}$ (with all parameters real). Clearly there are countless others, but characterizing the whole range of satisfactory $(a,b,c)$ values is beyond the scope of this paper. This analysis stands here to show how one can gain insight by looking at the output state of the neural network.}

To conclude our discussion of the bipartite case, we consider a family of entangled states that feature bound entanglement. Specifically, we consider the class of states introduced in Ref.~\cite{horodecki_bound_1999}, although we adopt the parametrization used in Ref.~\cite{mintert_measures_2005}. This family of 2-qutrit states exhibits bound entanglement, i.e. a PPT entangled region. The states are 
\begin{equation}\label{eq:minter}
    \rho_q = \dfrac{1}{21}
    \begin{pmatrix}
        2 & 0 & 0 & 0 & 2 & 0 & 0 & 0 & 2 \\
        0 & \beta_- & 0 & 0 & 0 & 0 & 0 & 0 & 0 \\
        0 & 0 & \beta_+ & 0 & 0 & 0 & 0 & 0 & 0 \\
        0 & 0 & 0 & \beta_+ & 0 & 0 & 0 & 0 & 0 \\
        2 & 0 & 0 & 0 & 2 & 0 & 0 & 0 & 2 \\
        0 & 0 & 0 & 0 & 0 & \beta_- & 0 & 0 & 0 \\
        0 & 0 & 0 & 0 & 0 & 0 & \beta_- & 0 & 0 \\
        0 & 0 & 0 & 0 & 0 & 0 & 0 & \beta_+ & 0 \\
        2 & 0 & 0 & 0 & 2 & 0 & 0 & 0 & 2
    \end{pmatrix},
\end{equation}
with $\beta_\pm = \dfrac{5}{2}\pm q$, and $q\in[-2.5,2.5]$, however, we only consider $q\in[0,2.5]$ since the negative $q$ regime gives the same states up to permutations. It is known that $\rho_q$ is separable for $q\in[0,0.5]$, is PPT entangled for $q\in(0.5,1.5]$ and is NPT entangled for $q\in(1.5,2.5]$. For several values of $q$ we train the neural network to approximate $\rho_\text{T} = \rho_q$, and display the results in Fig.~\ref{fig:iso_werner}. We can see that by explicitly constructing the separable decomposition, our results are not sensitive to whether the partial transpose is positive or negative, and the neural network approach successfully identifies the separable and entangled regions.

\begin{figure*}[t]
    \centering
    \includegraphics[width =\textwidth]{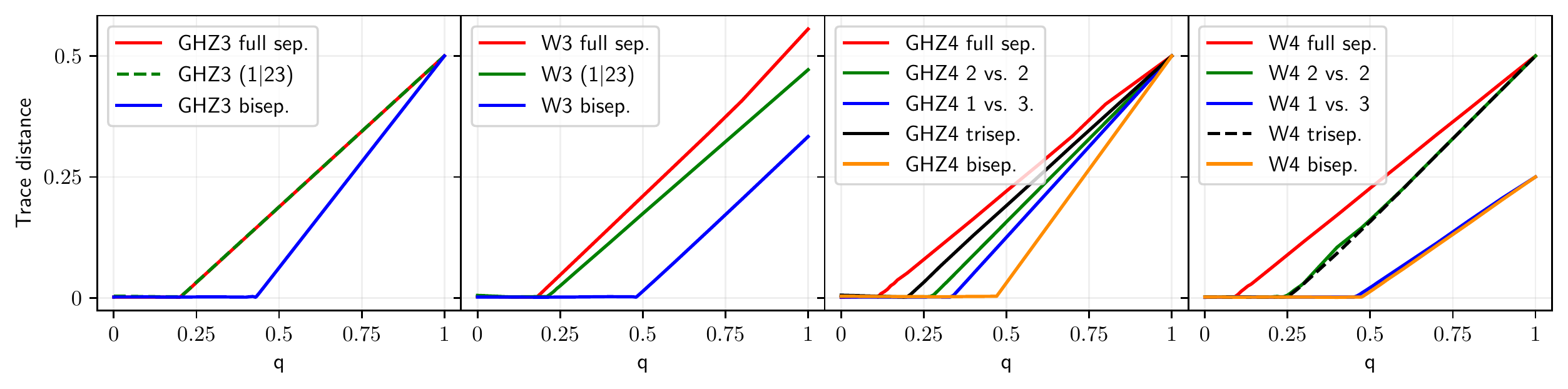}
    \caption{Neural network run on three and four-qubit GHZ and W states for different $q$ values, and different separable models. See Tables~\ref{table:GHZ_W_3} and \ref{table:GHZ_W_4} for the extracted threshold estimates. The 3-qubit GHZ full separability and (1|23) separability, as well as the 4-qubit W biseparability and 1 vs. 3 biseparability, and triseparability and 2 vs. 2 biseparability curves almost completely overlap, making them difficult to distinguish on the plot.}
    \label{fig:GHZ_W}
\end{figure*}

\subsection{Multipartite case}
We now consider three and four qubit multipartite states by examining the exemplary GHZ and W states, mixed with white noise. The GHZ state is
\begin{equation}
\ket{\text{GHZ}} = \frac{1}{\sqrt{2}} \left(\ket{00\dots0}+ \ket{11\dots1}\right),
\end{equation}
while the W state is
\begin{equation}
\ket{\text{W}} = \frac{1}{\sqrt{n}} \left((\ket{0\dots01}+ \ket{0\dots10} + \dots +\ket{10\dots0}\right).
\end{equation}
We mix both with the maximally mixed state as we did for the isotropic states in Eq.~(\ref{eq:isotropic}). For three qubits, we use the neural network to distinctly examine
\begin{enumerate}
\item full separability, as in Eq.~(\ref{eq:fully_separable}) ($n=3$),
\item biseparability with respect to a single partition (1$|$23), as in Eq.~(\ref{eq:biseparable_given_partition}),
\item biseparability, as in Eq.~(\ref{eq:biseparable}),
\end{enumerate}
and for the four qubits,
\begin{enumerate}
\item full separability, as in Eq.~(\ref{eq:fully_separable}) ($n=4$),
\item biseparability with respect to the partition (12$|$34), as in Eq.~(\ref{eq:biseparable_given_partition}),
\item biseparability with respect to the partition (1$|$234), as in Eq.~(\ref{eq:biseparable_given_partition}),
\item biseparability with respect to 2 vs. 2 partitions, i.e. as in Eq.~(\ref{eq:biseparable}), except all partitions $\mathcal{I}_k$ are constrained to have 2 parties,
\item biseparability with respect to 1 vs. 3 partitions, i.e. as in Eq.~(\ref{eq:biseparable}), except all partitions $\mathcal{I}_k$ are constrained to have 1 party (and thus the complements $\bar{\mathcal{I}}_k$ have 3 parties),
\item triseparability, as a generalization of Eq.~(\ref{eq:biseparable}), namely $\rho = \sum_{k} p_k \rho_{\mathcal{I}_k^1}^k \otimes \rho_{\mathcal{I}_k^2}^k \otimes \rho_{\mathcal{I}_k^3}^k$, with $(\mathcal{I}_k^1|\mathcal{I}_k^2|\mathcal{I}_k^3)$ a partitioning of $\mathcal{I}$ for each $k$, and
\item biseparability, as in Eq.~(\ref{eq:biseparable}).
\end{enumerate}
For the biseparable and triseparable cases, on a technical level, for each $k$ we ask the neural network to output density matrices for all possible partitions, i.e. for each $k$ it actually outputs 3 terms at a time for the 3-party case, and 6 terms for the 4-party case.


We present the results in Fig.~\ref{fig:GHZ_W}, except for 4-qubit separability with respect to a fixed partition, to not overcrowd the figure, however, note that those results are qualitatively similar. The consistent straight lines formed from independent runs give us confidence that the algorithm works well for approximately detecting the separability boundaries.

From Fig.~\ref{fig:GHZ_W}, we extract estimates of the separability bound by fitting linear curves to the data that is outside the seemingly flat separable region \textcolor{correction}{and close to the boundary}\footnote{Here the curves are close to linear, however, in general the curve could be nonlinear. In such cases a linear fit gives a good first-order estimate if it is fit to points near the boundary.}. \textcolor{correction}{Moreover, inspired from Ref.~\cite{shang_convex_2018}, we show how to obtain actual lower bounds on the separability thresholds in Appendix~\ref{app:thresholds}.} All results are given in Tables~\ref{table:GHZ_W_3} and \ref{table:GHZ_W_4}. With the flexibility of the current technique, we are able to quickly get estimates and bounds on the noise thresholds for many notions of separability, or alternatively, entanglement. These results can be improved by taking more points or running the algorithm multiple times. In cases where the exact threshold is known, our results are close to it. Where the boundary is not known to be exact, we can see how close it is to being tight. 
We observe that in these cases, in fact the analytic upper bounds seem to be close to, or in fact, optimal. Finally, we established estimates for many notions of separability, for which we did not find previous estimates or bounds in the literature. These were complemented by lower bounds provided by J.Shang and O. G\"uhne based on the method in Ref.~\cite{shang_convex_2018} in private communications.

\onecolumngrid
\begin{center}
\begin{table}[]
\caption{Three-party separability thresholds for the noisy GHZ and W states. Previously known bounds are all upper bounds. An asterisk (*) denotes that the bound is known to be exact. \textcolor{correction}{Our results are given with an estimation of the bound obtained with a linear fit close to the threshold and with a certified lower bound obtained as discussed in Appendix~\ref{app:thresholds}.}}
\label{table:GHZ_W_3}
\begin{tabular}{r||c|c|c|c|c|c}
             & \multicolumn{3}{c|}{3-qubit GHZ}                                               & \multicolumn{3}{c}{3-qubit W}                                                 \\
Separability & \multicolumn{1}{c|}{Estimate with linear fit} & \multicolumn{1}{c|}{\textcolor{correction}{Lower bound}}& \multicolumn{1}{c|}{Previous bound} & \multicolumn{1}{c|}{Estimate with linear fit}& \multicolumn{1}{c|}{\textcolor{correction}{Lower bound}} & \multicolumn{1}{c}{Previous bound} \\ \hline
Full sep. & 0.199 & 0.197 & 0.2*~\cite{dur_classification_2000} & 0.177 &0.177& 0.178*~\cite{chen_noise_2020}\\
1|23 sep. & 0.198 &0.198& 0.2* \cite{dur_classification_2000}& 0.206 &0.208&0.210~\cite{szalay_separability_2011}\\
Bisep.& 0.428 &0.423& 0.429*~\cite{guhne_separability_2010}& 0.478 &0.473& 0.479*\cite{jungnitsch_taming_2011}                                 
\end{tabular}
\end{table}

\begin{table}[]
\caption{Four-party separability thresholds for the noisy GHZ and W states.  \textcolor{correction}{Results are given as (i) an estimate of the bound obtained via a linear fit close to the threshold and (ii) a certified lower bound obtained from the method discussed in Appendix~\ref{app:thresholds}.} Bounds marked with a dagger ($\dagger$) are lower bounds provided by J.Shang and O. G\"uhne based on the method in Ref.~\cite{shang_convex_2018} in a private communication. All other previously known bounds are upper bounds. An asterisk (*) denotes cases where the bound is known to be exact.}
\label{table:GHZ_W_4}
\begin{tabular}{r||c|c|c|c|c|c}
             & \multicolumn{3}{c|}{4-qubit GHZ}                                               & \multicolumn{3}{c}{4-qubit W}                                                 \\
Separability & \multicolumn{1}{c|}{Estimate with linear fit} & \multicolumn{1}{c|}{\textcolor{correction}{Lower bound}}& \multicolumn{1}{c|}{Previous bound} & \multicolumn{1}{c|}{Estimate with linear fit}& \multicolumn{1}{c|}{\textcolor{correction}{Lower bound}} & \multicolumn{1}{c}{Previous bound} \\ \hline
Full sep.& 0.107&0.101& 0.111*~\cite{dur_classification_2000}& 0.086& 0.073 & 0.093*~\cite{chen_noise_2020}\\
(12|34) & 0.109 &0.109 & 0.111*~\cite{dur_classification_2000} & 0.109 &0.109 & 0.113$^\dagger$  \\
(1|234) & 0.108 &0.107 & 0.111*~\cite{dur_classification_2000} & 0.123 &0.124& 0.128$^\dagger$  \\
2 vs. 2 sep. & 0.271 & 0.267 & 0.269$^\dagger$ & 0.239 & 0.226 & 0.245$^\dagger$ \\
1 vs. 3 sep. & 0.333 & 0.328 & 0.327$^\dagger$& 0.452 & 0.435 & 0.447$^\dagger$ \\
Trisep.      & 0.198 & 0.194 &  0.198$^\dagger$  & 0.247 & 0.241& 0.243$^\dagger$\\
Bisep.       & 0.467 & 0.451 & 0.467*~\cite{jungnitsch_taming_2011} & 0.472 & 0.458 & 0.474~\cite{jungnitsch_taming_2011}  \\
\end{tabular}
\end{table}
\end{center}
\twocolumngrid

\section{Conclusion and outlook}
In summary, we have addressed the question of constructing the closest separable state to a given target state, by using a neural network as a compact model for separable states. We avoided the bottleneck of having to explicitly model many (up to $\prod_{j=1}^n d_j^2$) separable pure states in a decomposition by using a single neural network to represent them all. We demonstrated that by training the model independently on multiple states from a family, we can identify the separability boundary well. We did this for examples where the boundaries are known, PPT entangled states, as well as 3- and 4-party multi-qubit states. \textcolor{correction}{Additionally, we showed how analytical insight can be gained from the output of the algorithm. In the bipartite case, we partially characterized the closest separable state for the two-qubit Bell state, and conjectured relations for the distances in case of arbitrary dimension. In the multipartite case we showed how to obtain strict lower bounds on the noise threshold.}

The technique presented here opens up avenues for a variety of numeric applications in quantum foundations. In particular, for any task with reasonable Hilbert space sizes, it is possible to optimize over the set of separable states, as long as the loss function is differentiable. Among other potential applications, it can be especially helpful for obtaining (estimates or bounds on) entanglement measures, measures of robustness, separable ground state energies, and with minor modifications can be easily adapted to finding the closest classical state. Moreover, a particularly fruitful avenue for research could be focused on combining our approach with other generative neural network approaches to quantum state representations, namely ``quantum neural network states''~\cite{carleo_solving_2017,melko_restricted_2019}, particularly their extension to density matrices~\cite{yoshioka_constructing_2019,hartmann_neural-network_2019,nagy_variational_2019,vicentini_variational_2019}. Using such an ansatz for the separability problem has been examined in Ref.~\cite{harney_mixed_2021}. Such prospects of further developing the algorithms give the promise of exciting novel numerical tools for a broad range of tasks, both for numerical work and gaining analytic insight.

\section{Code availibility}
We have made sample code available at \url{www.github.com/Antoine0Girardin/Neural-network-for-separability-problem}.

\section{Acknowledgments}
We thank Pavel Sekatski for discussions, and Otfried Gühne and Jiangwei Shang for pointing out the work of Ref. \cite{shang_convex_2018} and for providing additional values for Table II. We acknowledge financial support from the Swiss National Science Foundation (project $2000021\_192244/1$ and NCCR QSIT). T.K. additionally acknowledges funding from the Swiss National Science Foundation Doc.Mobility grant (project P1GEP2\textunderscore199676).

\appendix
\section{Technical details of the utilized neural networks}\label{app:NN_details}
The main idea of how we use neural networks can be found in the maintext, while the implemented code can be found in the online repository provided. Here, we briefly describe some of the technical details and hyperparameters that we used.

As described in the maintext we use a feedforward neural network to represent a generic separable state of a fixed dimension and separability structure. We use a multilayer perceptron with rectified linear units as activations, except in the final layer where we use sigmoid activations. The outputs are normalized via a softmax function for the probability vectors, and by dividing by the 2-norm for the complex entries of the pure states. For the calculation in the maintext we employed a single hidden layer, with a width of 100, or 200 for more difficult calculations. The number of elements in the separable decomposition, $K$, is analytically upper bounded by $\prod_j d_j^2$, however in the implementation, typically $K = \prod_j d_j$ gives satisfactory results and allows for much quicker training. For training we use the Adadelta optimizer.

\section{Comparing with gradient descent}\label{app:SGD}
To see the advantage of using a neural network, we compare our algorithm with the naive optimization algorithm of gradient descent, for the simplest case of two qubits.

We parametrize the quantum state in a similar way as in Eq.~(\ref{eq:parametrized_bipartite}), i.e. the free parameters are the $K$ probabilities and the real and imaginary parts of the pure states composing the separable state according to Eq.~(\ref{eq:separable}), with $d=2, K=16$. The gradient descent algorithm varies these parameters to minimize the trace distance with respect to a target state, which we chose to be the Bell state, namely Eq.~(\ref{eq:max_ent_state}), with $d=2$. The gradient descent algorithm was run with an initial learning rate of 1, decreased by a factor of 0.98 each round for 250 rounds, and with a momentum factor of 0.2.

\begin{figure}[t]
    \centering
    \includegraphics[width =0.485\textwidth]{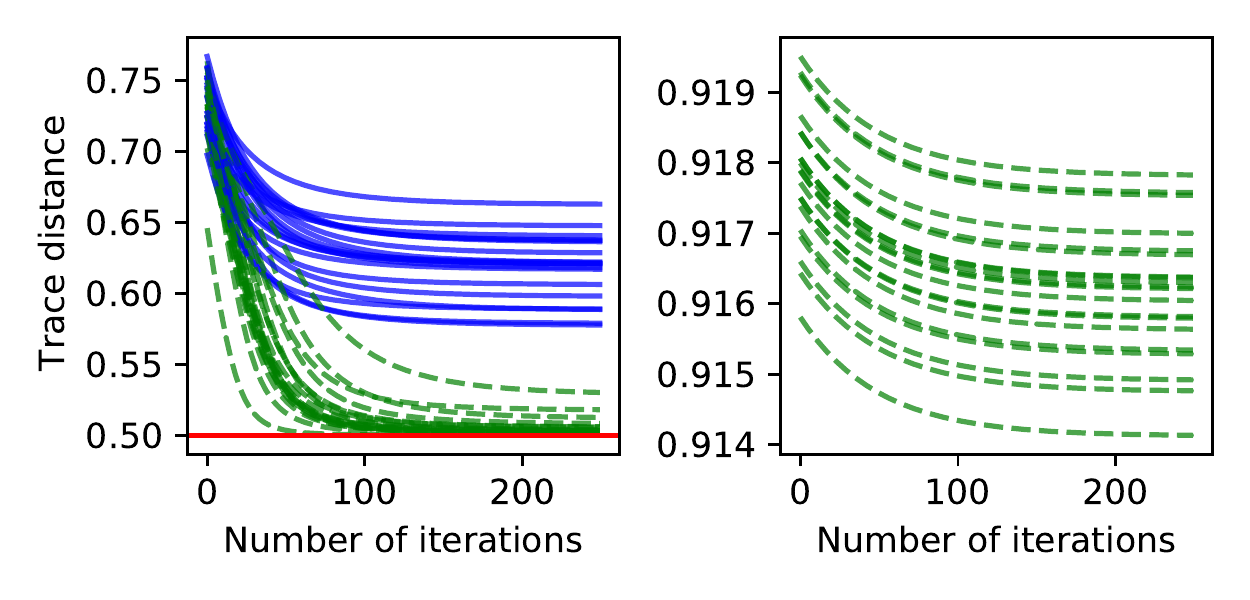}
    \caption{(Left) Trace distance from target Bell state versus iteration of gradient descent optimization for complex-valued (solid blue lines) and real-valued (dashed green lines) density matrices, for 20 independent runs each. The true minimum of 0.5 is depicted as a solid red line, which the neural network reaches up to at about 1E-7 precision; result taken from Fig.~\ref{fig:iso_werner}. (Right) Same plot for the real-valued gradient descent on the isotropic state, Eq.~(\ref{eq:isotropic}), for $d=5$, where the expected closest distance should be 0.8 (which the neural network approaches to 1E-6 precision).}
    \label{fig:AppA_1}
\end{figure}
Recall that the neural network, even with one layer, did not have any trouble finding the closest separable state with a trace distance of 0.5. However, as shown in the left panel of Fig.~\ref{fig:AppA_1}, we notice that already for this simple case the gradient descent technique has difficulties in finding the closest state. Somewhat surprisingly, if only real numbers are chosen to represent the state, the gradient descent technique performs better and converges to a good solution. Note that for higher dimensions, e.g. $d=5$, the real-valued gradient descent also has difficulties, as shown in the right panel of Fig.~\ref{fig:AppA_1}.

\section{Random states}\label{app:random}
\begin{figure}[t]
    \centering
    \includegraphics[width =0.49\textwidth]{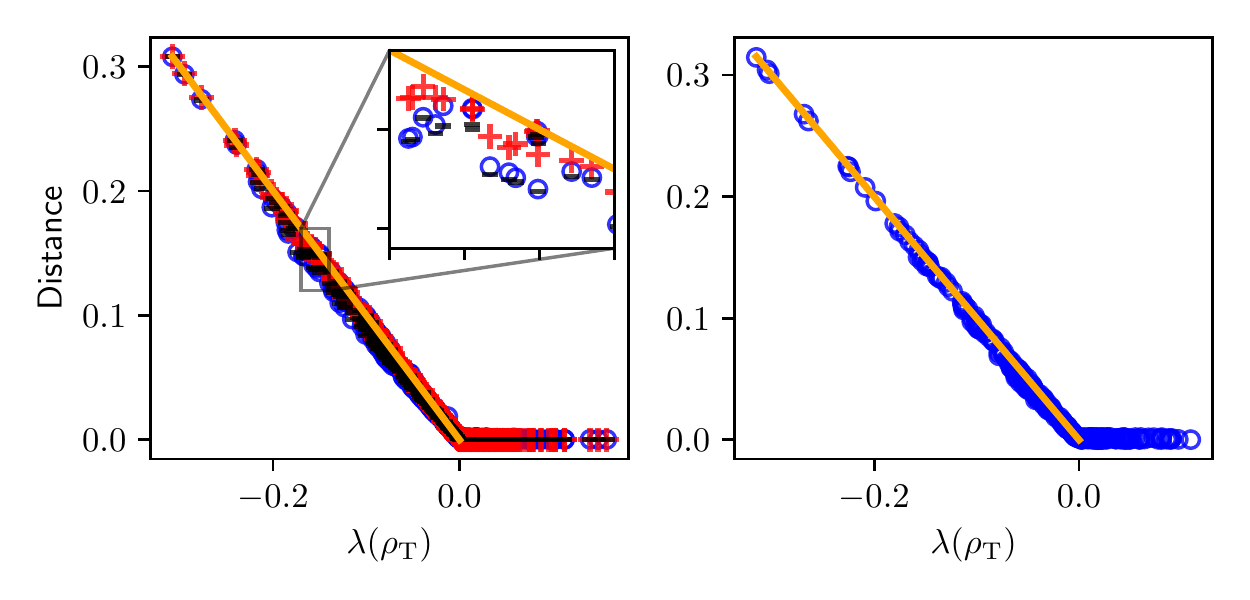}
    \caption{Neural network run on random two-qubit states, with the resulting distances scatter plotted as a function of the smallest eigenvalue of the partial transpose (blue circles).  Solid orange lines denote the line of slope -1. (Left) Trained on and plotted with trace distance for 400 samples. Additionally, the ansatz states' trace distances are depicted with red crosses, \textcolor{correction}{and the SDP results are plotted with black horizontal bars.} (Right) Trained on and plotted with Hilbert-Schmidt distance for 300 samples.}
    \label{fig:FigAppC_1}
\end{figure}
When benchmarking the method on random states, we noticed that there is a strong connection between the obtained trace distance of the closest separable state and the lowest eigenvalue of the partial transpose. In this appendix, we first show benchmark results for the method on random two-qubit states ($d_1=d_2=2$), where the PPT criteria clearly distinguishes entangled from separable states. We observe a strong correlation between the trace distance and Hilbert--Schmidt distance of the closest separable state and the smallest eigenvalue of the partial transpose of the state. \textcolor{correction}{We verify these results with a SDP to see how close to the optimal solution the neural network can get for two-qubits states.} Finally, we present an analytic ansatz of the closest separable state, based on the numerical results of the neural network and our intuition, which we numerically validate to be very close to the actual closest separable state.

In the two-qubit case the positive partial transpose criteria is a necessary and sufficient condition for separability. Thus, in Fig.~\ref{fig:FigAppC_1} we plot the distance to the closest separable state obtained by the neural network against the smallest eigenvalue of the partial transpose, which we will refer to as $\lambda$. Using the trace distance as a loss, we tested 400 random states with the trace distance as a loss function, and 300 with the Hilbert--Schmidt distance as the loss (the neural network was retrained 5 times for each state and the lowest distance was kept).

First, we observe that the neural network achieves close to zero distance in the separable regime for all states. Clearly it can not and should not reach zero distance for entangled states (i.e. on the left side of the figures, where $\lambda<0$). We observe a much stronger relation: in fact the Hilbert--Schmidt distances of the closest separable state seem to line up on a line with slope $-1$, while the trace distance results seem to be below this. We formulate these two observations; namely in the entangled regime, for $\lambda(\rho_\text{T})<0$,
\begin{align}
    D_\text{HS}(\rho_\text{T}, \rho_\text{CSS;HS}) &\leq -\lambda(\rho_\text{T}),\\
    D_\text{Tr}(\rho_\text{T}, \rho_\text{CSS;Tr}) & \leq -\lambda (\rho_\text{T}),
\end{align}
where we explicitly denoted which distance was minimized in the subscript of $\rho_\text{CSS}$.

\textcolor{correction}{
It is possible to use a SDP to find the closest separable state with respect to the trace distance by using the PPT criteria.}

\textcolor{correction}{The SDP has a dual form and can be expressed as \cite{Fazel01} }
\textcolor{correction}{
	\begin{equation}
		\begin{split}
		&\text{minimize } Tr Y + Tr Z \\
		&\text{subject to }
		\begin{pmatrix}
			Y	& X   \\
			X^\dagger			& Z  
		\end{pmatrix} \geq 0	
		\end{split}
	\end{equation}
}
\textcolor{correction}{With $Y$ and $Z$ Hermitian matrices, $X=\rho_T-\rho_{CSS}$, $\rho_T$ being the target state and $\rho_{CSS}$ the separable state, $\rho_{CSS}\geq 0$, $tr(\rho_{CSS})=1$, $\rho_{CSS}^{\Gamma}\geq 0$. $\rho_{CSS}^{\Gamma}$ is the partial transpose of $\rho_{CSS}$.}

\textcolor{correction}{The result of the SDP for all random states are plotted in Fig.~\ref{fig:FigAppC_1}. }

Finally, we provide an ansatz for the closest separable state with respect to the trace distance. Intuitively, we set the smallest eigenvalue of the partial \textcolor{correction}{transpose} to be 0 instead of negative, and adjust the others such that the trace remains unchanged.
\begin{theorem*}
Let $\rho_\text{T}$ be an entangled state whose partial transpose has an eigendecomposition of $U D U^\dag$, with $D=\text{diag}(\lambda_1,\lambda_2,\lambda_3,\lambda_4)$, where $\lambda_1$ is the smallest eigenvalue (i.e. $\lambda_1 \equiv \lambda(\rho_\text{T})$). Then let our ansatz of the closest separable state be $\rho' = (U D' U^\dag)^\Gamma$ with $D' = \text{diag}(0, \lambda_2+\frac{\lambda_1}{3}, \lambda_3+\frac{\lambda_1}{3}, \lambda_4+\frac{\lambda_1}{3})$ and $X^\Gamma$ denoting the partial transpose of $X$. If $\rho'$ is a valid density matrix then
\begin{align}
    D_\text{Tr}(\rho_\text{T}, \rho') \leq -\lambda (\rho_\text{T}).
\end{align}
\end{theorem*}
Before proceeding to the proof, note that $\rho'$ is only actually a separable density matrix if $\lambda_2+\frac{\lambda_1}{3}>0$. However, only about $0.1\%$ of random states have a $\rho'$ approximation which is not a valid separable density matrix. The trace distances of the approximations of the 400 random states examined previously are depicted in Fig.~\ref{fig:FigAppC_1}.

\begin{proof}
Recall that $D_\text{Tr}(\rho_\text{T}, \rho') = \frac{1}{2} \sum_i |\mu_i|$, where $\{\mu_i\}_i$ is the set of eigenvalues of the difference. As a first step let us examine this difference.
\begin{align*}
    \rho_\text{T}-\rho' &= \left(U (D - D')U^\dag\right)^\Gamma =\\
    &= (U\begin{pmatrix}
    \lambda_1 & 0& 0& 0\\
    0 & -\lambda_1/3& 0& 0\\
    0 & 0& -\lambda_1/3& 0\\
    0 & 0& 0& -\lambda_1/3
    \end{pmatrix}U^\dag)^\Gamma =\\
    &= \lambda_1 \left( \frac{4}{3}U E_{11} U^\dag - \mathbb{I}/3\right)^\Gamma =\\
    &= \lambda_1 \left( \frac{4}{3} u u^\dag  - \mathbb{I}/3\right)^\Gamma,
\end{align*}
where $E_{11}$ is the matrix with a single nonzero entry in its first position, and thus $u$ is the first column of $U$.

To prove the theorem we must show that no matter what $u$ appears in the decomposition, the trace distance is bounded, namely that
\begin{align}
    \max_{u} D_\text{Tr}(\rho_T,\rho') \leq -\lambda_1,
\end{align}
which, after canceling out $-\lambda_1$, reads explicitly as
\begin{align}\label{eq:appC_ineq}
    \max_{u} \frac{1}{2} \sum_i \left| \text{e.v.}_i  \left( \frac{4}{3} (u u^\dag)^\Gamma  - \mathbb{I}/3\right) \right| \leq 1,
\end{align}
where we have used the notation $\text{e.v.}_i$ for the $i$-th eigenvalue. Using that the identity matrix is jointly diagonalizable with $uu^\dag$, the left-hand side becomes
\begin{align}\label{eq:appC_expression}
    \max_{u} \frac{1}{6} \sum_i \left| 4 \nu_i  - 1 \right|,
\end{align}
where $\{\nu_i\}_{i=1}^4$ are the eigenvalues of $(u u^\dag)^\Gamma$ in non-decreasing order. Notice that the partial transpose preserves the trace, so $\sum_i\nu_i= 1$, since $u$ is the (unit-length) first column of a unitary matrix. 
So essentially, we must maximize (\ref{eq:appC_expression}) by distributing 1 among the four eigenvalues $\nu_i$. Due to the absolute value, the value $\frac{1}{4}$ becomes a divider: eigenvalues below it should be as small as possible, while eigenvalues above it should be as large as possible. So we split the eigenvalues into two parts
\begin{align*}
    S_{\leq \frac{1}{4}} &:= \{\nu_i|\nu_i \leq \frac{1}{4} \}, \\
    S_{> \frac{1}{4}} &:= \{\nu_i|\nu_i > \frac{1}{4} \}.
\end{align*}
Case-by-case we give an upper bound for Expression (\ref{eq:appC_expression}), based on the number of $\nu_i$ in $S_{\leq\frac{1}{4}}$. For any eigenvalues appearing in $S_{>\frac{1}{4}}$ the absolute value just disappears when upper bounding Expression (\ref{eq:appC_expression}). So if $S_{\leq\frac{1}{4}} = \emptyset$, then 
\begin{align}
\max_u \sum_i|4\nu_i -1| \leq \sum_i (4\nu_i -1)= 4 \sum_i(\nu_i)- 4 = 0.
\end{align}
If we have $S_{\leq\frac{1}{4}} = \{\nu_1\}$, then the best we can do is push down $\nu_1$ to be as negative as possible, so the other eigenvalues can jointly be larger ($\nu_2+\nu_3+\nu_4 = 1 + |\nu_1|$ if $\nu_1<0$). Additionally, it is known that there can be at most one negative eigenvalue of the partial transpose\cite{rana_negative_2013,johnston_family_2010}, and that all eigenvalues are larger than $-1/2$~\cite{rana_negative_2013}, i.e.
\begin{align}
    -\frac{1}{2} &\leq \nu_1, \\
      0 &\leq \nu_2. \label{eq:appC_partialtrace_eigenvalues}
\end{align}
Using the first, and that the eigenvalues sum to 1, we see that
\begin{align}
\max_u \sum_{i=1}^4|4\nu_i -1| &= \max_u \sum_{i=2}^4 (4\nu_i -1) -(4\nu_1-1) = \\
&= \max_u 4\left(\sum_{i=2}^4 \nu_i\right) - 4\nu_1 - 2 \leq\\
&\leq 4(1-\nu_1) - 4\nu_1 -2 =\\
&= 2 - 8\nu_1 \leq 2 + \frac{8}{2} = 6.
\end{align}
Finally, note that it does not make sense to add more eigenvalues to $S_{\leq\frac{1}{4}}$, since if $S_{\leq\frac{1}{4}} = \{\nu_1,\nu_2\}$, then by Ineq.~(\ref{eq:appC_partialtrace_eigenvalues}), namely that $0\leq \nu_2$, we cannot increase the weight of $S_{>\frac{1}{4}}$, i.e. $\nu_3+\nu_4 = 1+|\nu_1|-\nu_2$. So essentially we are in the same position as when $S_{\leq\frac{1}{4}} = \{\nu_1\}$, and thus the upper bound is 6. Placing this back in Expression(\ref{eq:appC_expression}), or Ineq.~(\ref{eq:appC_ineq}), we see that the theorem is proven.
\end{proof}

\textcolor{correction}{
\section{Lower bounds on separability thresholds}\label{app:thresholds}
To be able to give separability thresholds and not only approximation with linear plots, we follow a procedure similar that the one used in Ref.~\cite{shang_convex_2018}. The core idea is to prove the separability of a target state by showing that it can be written as the convex combination of two separable states. One of the two states is separable because it is generated by the neural network ($\rho_{\text{CSS}}$ in the following). The other is separable because it lies within the separability ball around the fully mixed state ($\rho_x$).
}

\begin{figure}[t]
    \centering
    \includegraphics[width =0.4\textwidth]{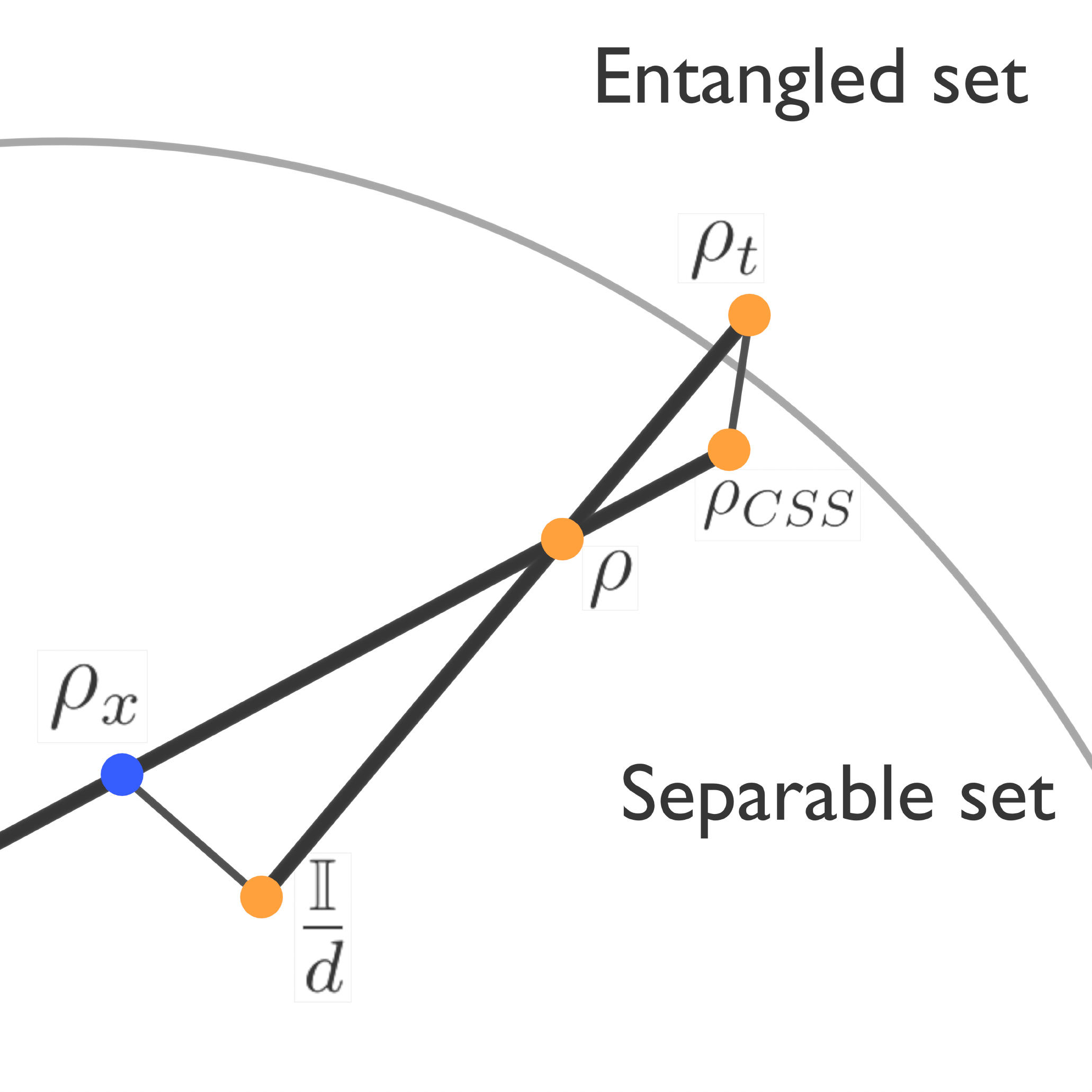}
    \caption{Schema of the method to get certification on lower bounds}
    \label{fig:FigAppD}
\end{figure}

\textcolor{correction}{
In more detail, we start from a state $\rho$ we want to prove separable, and build a new state $\rho_t=(1+\epsilon)\rho - \epsilon\frac{\mathbb{I}}{d}$, where $\epsilon>0$ is an arbitrary small number such that $D_{\text{Tr}}(\rho_t,\rho)$ is at least larger than the neural network's precision. That new state $\rho_t$ will be further away from the fully mixed state. Then we use the neural network to obtain an approximately closest separable state to $\rho_t$, which we call $\rho_{\text{CSS}}$. Finally, we try to find a separable state $\rho_x = (1+\epsilon')\rho - \epsilon'\rho_{CSS}$ by scanning several values of $\epsilon'$. We can use the condition $tr(\rho_x^2)\leq\frac{1}{2^N-\alpha^2}$ with $\alpha^2=\frac{2^N}{\frac{17}{2}3^{N-3}+1}$ and $N$ the number of parties, from Ref.~\cite{barreiro_experimental_2010} to show that $\rho_x$ is fully separable. The condition $tr(\rho_x^2)\leq\frac{1}{d-1}$ with $d$ the dimension of the Hilbert space from Ref.~\cite{gurvits02} can be used to show biseparability. If the condition is satisfied, the separability of $\rho$ follows from the convexity of the set of separable states.
} 


	%
	
\end{document}